\documentclass[preprint,showpacs,preprintnumbers,amsmath,amssymb]{revtex4}
\usepackage{graphicx}

\begin{document}
\def \ee {\varepsilon}
\thispagestyle{empty}
\title{
Application of the proximity force approximation
to gravitational and Yukawa-type forces
}

\author{R.~S.~Decca,${}^1$ E.~Fischbach,${}^2$
G.~L.~Klimchitskaya,${}^3$
D.~E.~Krause,${}^{4,2}$ D.~L\'{o}pez,${}^5$
and V.~M.~Mostepanenko${}^6$
}

\affiliation{
${}^1$Department of Physics, Indiana University-Purdue
University Indianapolis, Indianapolis, Indiana 46202, USA\\
${}^2$Department of Physics, Purdue University, West Lafayette, Indiana
47907, USA\\
${}^3$North-West Technical University, Millionnaya St. 5, St.Petersburg,
191065, Russia \\
${}^4$Physics Department, Wabash College, Crawfordsville, Indiana 47933,
USA\\
${}^5$Center for Nanoscale Materials, Argonne National Laboratory,
Argonne, Illinois 60439, USA \\
${}^6$Noncommercial Partnership
``Scientific Instruments'',  Tverskaya St. 11, Moscow,  103905, Russia
}

\begin{abstract}
We apply the proximity force approximation, which is
widely used for the calculation of the Casimir force
between bodies with nonplanar boundary surfaces,
to gravitational and Yukawa-type interactions. It is
shown that for the gravitational force in a sphere-plate
configuration the general formulation of the proximity
force approximation is well applicable. For a
Yukawa-type interaction we demonstrate the validity of
both the general formulation of the proximity
force approximation, and a simple
mapping between the sphere-plate and plate-plate
configurations. The claims to the contrary in some recent
literature are thus incorrect.
Our results justify the constraints on the parameters
of non-Newtonian gravity previously obtained from the indirect
dynamic measurements of the Casimir pressure.
\end{abstract}

\maketitle

\section{Introduction}

During the last few years, much attention has been devoted to
experimental
tests of Newton's gravitational law at short separations,
and to the search for possible corrections to it (see, e.g., monograph
\cite{1} and review \cite{2}). These corrections arise from the
exchange of massless or light elementary particles predicted in
many models containing a spontaneously or weakly dynamically broken
symmetry. For example, the exchange of light bosons, such as
scalar axions, graviphotons, dilatons and moduli (see, e.g.,
Refs.~\cite{3,4}) generates a Yukawa-type correction to Newtonian
gravity. An almost identical correction is predicted
 in extra-dimensional physics with compact extra dimensions and
 low-energy compactification scale of order 1\,TeV \cite{5,6,7,8}.

 Constraints on corrections to Newton's gravitational law at
 short separations can be obtained from precise force
 measurements between macrobodies. At separations greater than
 $10^{-5}\,$m the dominant force between electrically neutral
 test bodies is gravity, and recent
 new strong constraints on the Yukawa-type interaction in
 sub-millimeter range have been obtained in gravity experiments
 \cite{9,10,11,12,13,14,15}. In the sub-micrometer range the
 dominant forces are the Casimir and van der Waals
 forces \cite{16}. Here the strongest constraints on Yukawa-type
 corrections to Newtonian gravity were obtained from
 measurements of the Casimir force
\cite{17,18,19,20,21,22,23,24,25}.

All precise measurements of the Casimir force which were used
to obtain constraints on non-Newtonian gravity were performed in
a sphere-plate configuration, rather than in a plate-plate
configuration. The reason is that it is hard to
ensure that two plates are parallel at separations $a$ below
$a=1\,\mu$m
with sufficient precision. The Casimir force
in a sphere-plate configuration $F_C(a)$ can be obtained from
the Casimir energy per unit area in a configuration of two
parallel plates, $E_C(a)$, by means of
\begin{equation}
F_C(a)=2\pi R E_C(a),
\label{eq1}
\end{equation}
\noindent
where $R$ is the sphere radius.
This equation represents the application of the general
approximate  method, the proximity force approximation (PFA)
\cite{26}, for this geometry.
As was demonstrated recently both
theoretically \cite{28,29,30} and experimentally \cite{31},
the PFA provides a very precise determination of the Casimir force
under the condition that $a\ll R$. The relative corrections
to the value of the force calculated using Eq.~(\ref{eq1})
were shown to be less than $a/R$, and for typical $a$ and $R$
used in experiments this is only 0.1\% of the Casimir
force. In the dynamic measurement regime, the PFA gives
the possibility to determine the Casimir pressure between
two parallel plates from experimental data for the
gradient of the Casimir force in a sphere-plate configuration.

It has been claimed recently \cite{32} that the PFA ``does not
hold for forces of volumetric character such as the
gravitational force or its hypothetical short-range relatives".
On this basis the limits on the Yukawa force obtained from the
dynamic measurements of the Casimir pressure between two
parallel plates using a sphere-plate configuration, as
in \cite{21} and some other experiments, were called
``invalid". Below we apply the PFA to both gravitational
and Yukawa-type forces and demonstrate that the results that are
obtained agree nicely with the results of direct
calculations that do not use the PFA.
Statements to the contrary made in Ref.~\cite{32}
with respect to gravitational force are explained by a
confusion with different formulations of the PFA.
With respect to the Yukawa-type interaction these
statements are shown to be incorrect.

The paper is organized as follows. In Sec.~II we present different
formulations of the proximity force approximation and apply them
to the gravitational interaction. Section~III demonstrates
that the proximity force approximation is readily applicable
to Yukawa-type hypothetical forces.
Section IV establishes the applicability of the PFA to Yukawa
forces in dynamic experiments constituting an indirect measurement of
the Casimir pressure between two parallel plates using the
sphere-plate configuration.
In Sec.~V we present our conclusions and discussion.

\section{The proximity force approximation and
Newtonian gravity}

In the most general formulation of the PFA \cite{26},
the $z$-component of the force acting between two arbitrarily
shaped bodies $V_1$ and $V_2$ can be approximately
represented as a sum of forces between plane parallel surface
elements $dx\,dy$
\begin{equation}
F_z(a)=\int\!\!\int_{\sigma}dx\,dy\,P\bigl(x,y,z(x,y)\bigr).
\label{eq2}
\end{equation}
\noindent
Here, $P\bigl(x,y,z(x,y)\bigr)$ is the known pressure for a
configuration of two parallel plates,
$z(x,y)=z_2(x,y)-z_1(x,y)$, where $z_2(x,y)>z_1(x,y)$
are the surfaces of the bodies arranged against each other
in an appropriate coordinate system, $\sigma$ is the part
of the $(x,y)$-plane, where both surfaces are defined,
and $a$ is the smallest value of $z(x,y)$.

In Ref.~\cite{27} Eq.~(\ref{eq2}) was applied to the case
of smooth interacting surfaces having a single point
$x=y=0$ where the width of the gap $z(x,y)$ reaches
a minimum. It was also assumed that the characteristic size
of the upper body in the $z$-direction extends beyond the
range of $z$ where the pressure $P\bigl(x,y,z(x,y)\bigr)$
drops to zero,
so that the integral becomes independent of its upper limit.
Under these conditions Eq.~(\ref{eq2}) yielded \cite{27}
\begin{equation}
F_z(a)=2\pi\bar{R}E(a),
\label{eq3}
\end{equation}
\noindent
where $E(a)$ is the energy per unit area in the configuration
of two parallel plates of infinite area, interacting via
the same force as the bodies $V_1$ and $V_2$,
$\bar{R}=\sqrt{R_xR_y}$ and $R_x,\>R_y$ are the principal
radii of curvature at the point (0,0).
For a sphere at short separations above a plate ($a\ll R$)
interacting via the Casimir force, Eq.~(\ref{eq3})
transforms into Eq.~(\ref{eq1}).

We next check the applicability of the PFA for the
calculation of the gravitational force acting between a
large plane plate of mass density $\rho_1$ and thickness
$D_1$ and a spherical ball (sphere) of radius $R$ and
density $\rho_2$ spaced at a height $a$ above this
plate (see Fig.~1). The exact calculation of the
gravitational force in this configuration is straightforward.
For this purpose we first consider the point-like
mass $m_2$ at a point $\mbox{\boldmath$r$}_2$ belonging
to the sphere interacting via Newton's force with any
point-like
mass $m_1$ at a point $\mbox{\boldmath$r$}_1$ belonging
to the plate
\begin{equation}
\mbox{\boldmath$F$}=-Gm_1m_2\frac{\mbox{\boldmath$r$}_2-
\mbox{\boldmath$r$}_1}{|\mbox{\boldmath$r$}_2-
\mbox{\boldmath$r$}_1|^3},
\label{eq4}
\end{equation}
\noindent
where $G$ is the gravitational constant.
The $z$-component of the force acting between the mass
$m_2$ and the plate is obtained by integration of
(\ref{eq4}) over the plate volume
\begin{eqnarray}
F_{z,{\rm gr}}^{m_2}&=&
-2\pi Gm_2\rho_1
\int_{-D_1}^{0}dz_1\int_{0}^{\infty}r\,dr
\frac{z_2-z_1}{\bigl[r^2+(z_2-z_1)^2\bigr]^{3/2}}
\nonumber \\
&=&-2\pi Gm_2\rho_1D_1,
\label{eq5}
\end{eqnarray}\noindent
where the $(x,y)$-plane coincides with the upper boundary
surface of the plate, and $r^2=x^2+y^2$.

Note that the force (\ref{eq5}) does not depend on
$z_2$. Because of this the integration over the
volume of the sphere reduces to the multiplication
by its volume. Thus the gravitational force between
a sphere and a plate is given by
\begin{equation}
F_{z,{\rm gr}}^{sp}=-\frac{8\pi^2}{3}
G\rho_1\rho_2D_1R^3.
\label{eq6}
\end{equation}
\noindent
From Eq.~(\ref{eq5}) it is also easy to obtain the
exact expression for the gravitational pressure between the
two parallel plates of infinite area, lower and upper, with
parameters $\rho_1,\> D_1$ and $\rho_2,\> D_2$,
respectively
\begin{equation}
P_{{\rm gr}}=-2\pi
G\rho_1\rho_2D_1D_2.
\label{eq7}
\end{equation}
\noindent
We emphasize that this pressure does not depend on the
separation between the plates.

Next we apply the most general formulation (\ref{eq2}) of the PFA
to calculate the gravitational force in the sphere-plate
configuration. The pressure $P\bigl(x,y,z(x,y)\bigr)$ in
Eq.~(\ref{eq2}) is given by Eq.~(\ref{eq7}) where in
accordance with Fig.~1 $D_2$ should be replaced by
\begin{equation}
D_2(x,y)=2\sqrt{R^2-x^2-y^2}=2\sqrt{R^2-r^2}.
\label{eq8}
\end{equation}
\noindent
Substituting Eqs.~(\ref{eq7}) and (\ref{eq8}) in
Eq.~(\ref{eq2}), we obtain the PFA result for the gravitational
force acting between a sphere and a plate
\begin{eqnarray}
F_{z,{\rm gr}}^{sp}&=&-8\pi^2G\rho_1\rho_2D_1
\int_{0}^{R}r\,dr\,\sqrt{R^2-r^2}
\nonumber \\
&=&-\frac{8\pi^2}{3}G\rho_1\rho_2D_1R^3,
\label{eq9}
\end{eqnarray}
\noindent
which is identical to the exact result (\ref{eq6}).
The fact that in a sphere-plate configuration the PFA leads
to the exact result is explained by the additivity of the
gravitational interaction. It is important to keep in mind,
however, that Eq.~(\ref{eq9}) cannot be obtained using
the second, simplified, formulation of the PFA given
by Eq.~(\ref{eq3}). The reason is that $P_{\rm gr}$ in
Eq.~(\ref{eq7}) does not drop to zero within the volume of
a sphere which makes Eq.~(\ref{eq3}) inapplicable.

\section{The proximity force approximation and Yukawa-type
interaction}

We next consider a Yukawa-type interaction potential
between the point-like material particles
\begin{equation}
V_{\rm Yu}(|\mbox{\boldmath$r$}_2-\mbox{\boldmath$r$}_1|)=
-Gm_1m_2\alpha\frac{1}{|\mbox{\boldmath$r$}_2-
\mbox{\boldmath$r$}_1|}
{\rm e}^{-|\mbox{\scriptsize{\boldmath$r$}}_2-
\mbox{\scriptsize{\boldmath$r$}}_1|/\lambda},
\label{eq10}
\end{equation}
\noindent
where $\alpha$ is the dimensionless interaction strength
relative to gravity, and $\lambda$ is the interaction
range, and we calculate the respective force acting between
a sphere and a plate. Keeping in mind that this
interaction is very weak, and is not of electromagnetic
origin, it can be considered as additive to a very high
degree of accuracy. We begin with the calculation of the
Yukawa force acting between a sphere and a plate by directly
performing the additive summation of potentials
(\ref{eq10}).

By considering a particle $m_2$ at a height $z$ above a large
plate with thickness $D_1$ and density $\rho_1$, we obtain
the Yukawa energy of their interaction
by integration of Eq.~(\ref{eq10})
over the volume of the plate
\begin{equation}
E_{\rm Yu}^{m_2}(z)=-2\pi Gm_2\rho_1\alpha\lambda^2
{\rm e}^{-z/\lambda}\bigl(1-{\rm e}^{-D_1/\lambda}\bigr).
\label{eq10a}
\end{equation}
\noindent
Calculating the negative derivative of this expression with
respect to $z$, one finds the $z$-component of the Yukawa
force acting between a particle and a plate
\begin{equation}
F_{z,{\rm Yu}}^{m_2}(z)=-2\pi Gm_2\rho_1\alpha\lambda
{\rm e}^{-z/\lambda}\bigl(1-{\rm e}^{-D_1/\lambda}\bigr).
\label{eq11}
\end{equation}
\noindent
Integrating this equation over the volume of the sphere
(see Fig.~1), one obtains the desired result for the
Yukawa force in a sphere-plate configuration
\begin{eqnarray}
F_{z,{\rm Yu}}^{sp}(a)&=&-2\pi^2G\rho_1\rho_2\alpha\lambda
\bigl(1-{\rm e}^{-D_1/\lambda}\bigr)
\nonumber \\
&&~~~~~~~~~~~~~
\times\int_{a}^{2R+a}dz\bigl[R^2-(R+a-z)^2\bigr]\,
{\rm e}^{-z/\lambda}
\nonumber \\
&=&-4\pi^2G\rho_1\rho_2\alpha\lambda^3
\bigl(1-{\rm e}^{-D_1/\lambda}\bigr)R{\rm e}^{-a/\lambda}
\nonumber \\
&&~~~~~~~~~~~~~
\times\left(1-\frac{\lambda}{R}+{\rm e}^{-2R/\lambda}+
\frac{\lambda}{R}\,{\rm e}^{-2R/\lambda}\right).
\label{eq12}
\end{eqnarray}
\noindent
{}From the integration of Eq.~(\ref{eq11}) over the volume of
the upper plate one can also obtain the expression for the
Yukawa pressure between the two parallel plates of infinite
area spaced at a separation $a$ with the parameters
$\rho_1,\> D_1$ (the lower plate) and
$\rho_2,\> D_2$ (the upper plate)
\begin{equation}
P_{\rm Yu}(a)=-2\pi G\rho_1\rho_2\alpha\lambda^2
{\rm e}^{-a/\lambda}
\bigl(1-{\rm e}^{-D_1/\lambda}\bigr)
\bigl(1-{\rm e}^{-D_2/\lambda}\bigr).
\label{eq13}
\end{equation}
\noindent
In a similar way, integrating Eq.~(\ref{eq10a}) over
the volume of an upper plate we arrive at the Yukawa
energy per unit area in the configuration of two
parallel plates
\begin{equation}
E_{\rm Yu}(a)=-2\pi G\rho_1\rho_2\alpha\lambda^3
{\rm e}^{-a/\lambda}
\bigl(1-{\rm e}^{-D_1/\lambda}\bigr)
\bigl(1-{\rm e}^{-D_2/\lambda}\bigr).
\label{eq14}
\end{equation}

We now turn to the application of the PFA to the Yukawa
interaction starting from the most general formulation
of PFA given by Eq.~(\ref{eq2}). For this purpose we
replace $a$ in Eq.~(\ref{eq13}) by
$R+a-\sqrt{R^2-r^2}$ (which is the separation
between the plane plates in accordance with Fig.~1) and
choose $D_2$ as in Eq.~(\ref{eq8}). Substituting the
resulting expression for $P_{\rm Yu}\bigl(x,y,z(x,y)\bigr)$
in Eq.~(\ref{eq2}) and performing the integration one
finds the PFA result for the Yukawa interaction in
a sphere-plate configuration
\begin{eqnarray}
F_{z,{\rm Yu}}^{sp}(a)&=&-4\pi^2G\rho_1\rho_2\alpha\lambda^2
\bigl(1-{\rm e}^{-D_1/\lambda}\bigr)
\nonumber \\
&&~~~~
\times\int_{0}^{R}r\,dr\,{\rm e}^{-(R+a-\sqrt{R^2-r^2})/\lambda}\,
\bigl(1-{\rm e}^{-2\sqrt{R^2-r^2}/\lambda}\bigr)
\nonumber \\
&=&-4\pi^2G\rho_1\rho_2\alpha\lambda^3
\bigl(1-{\rm e}^{-D_1/\lambda}\bigr)R{\rm e}^{-a/\lambda}
\nonumber \\
&&~~~~
\times\left(1-\frac{\lambda}{R}+{\rm e}^{-2R/\lambda}+
\frac{\lambda}{R}\,{\rm e}^{-2R/\lambda}\right).
\label{eq15}
\end{eqnarray}
\noindent
It is seen that Eq.~(\ref{eq15}) is identical to Eq.~(\ref{eq12}).
Thus, for the Yukawa interaction the most general formulation
of the PFA leads to exactly the same result as the additive
summation of potentials (\ref{eq10}).

Now we consider a more subtle situation with the simplified
formulation of the PFA given by Eq.~(\ref{eq3}).
{}From the comparison of Eqs.~(\ref{eq14}) and (\ref{eq15})
it is seen that $F_{z,{\rm Yu}}^{sp}(a)$ is not obtained as
$E_{\rm Yu}(a)$ times $2\pi R$, as is required by Eq.~(\ref{eq3}).
However, in the case of a sufficiently thick upper plate
($D_2\gg\lambda$) and large sphere ($R\gg\lambda$)
one arrives from Eqs.~(\ref{eq14}), (\ref{eq15}) at
\begin{eqnarray}
F_{z,{\rm Yu}}^{sp}(a)&=&-4\pi^2G\rho_1\rho_2\alpha\lambda^3
\bigl(1-{\rm e}^{-D_1/\lambda}\bigr)R{\rm e}^{-a/\lambda}
\nonumber \\
&=&2\pi RE_{\rm Yu}(a)
\label{eq16}
\end{eqnarray}
\noindent
in accordance with the simplified formulation of the PFA
in Eq.~(\ref{eq3}). Thus, the simplified formulation of
the PFA works well for a Yukawa-type interaction
providing the mapping between the configurations of a large
sphere ($R\gg\lambda$) above a plate and the same plate
parallel to another (thick) plate.

\section{Layered structures used in dynamic experiments}

We next consider the application of the above results to the
configuration of experiments in Refs.~\cite{20,21,22,23}.
In these experiments, the Casimir pressure between two
parallel plates was measured indirectly in the dynamic
mode using experimental data for the gradient of
the Casimir force, acting between a sphere and a
plate, and the PFA.
The extent of agreement of the resulting data for
the Casimir pressure with theory was used to obtain
constraints on a Yukawa-type hypothetical interaction.
In so doing the expression for the Yukawa
pressure in the configuration of two parallel plates
has been used. To justify this calculation
procedure we require that the PFA provide a similar
mapping between the sphere-plate and plate-plate
configurations for both Casimir and Yukawa-type forces.
The latter was questioned in Ref.~\cite{32}, but
shown to be correct in Eq.~(\ref{eq16}) for the case
of a homogeneous plate ($D_2\gg\lambda$) and sphere
($R\gg\lambda$) with densities $\rho_1$ and $\rho_2$,
respectively.

In the experiments of Refs.~\cite{20,21,22,23}, however,
the plate and the sphere were not homogeneous but
covered with thin metallic layers with thicknesses
$\Delta_1^{\!\prime},\>\Delta_1^{\!\prime\prime}$
and densities $\rho_1^{\prime},\>\rho_1^{\prime\prime}$
(for the plate) and thicknesses
$\Delta_2^{\!\prime},\>\Delta_2^{\!\prime\prime}$
and densities $\rho_2^{\prime},\>\rho_2^{\prime\prime}$
(for the sphere). Below we check that the mapping provided by
the PFA for the configurations of sphere-plate and plate-plate
in the case of Yukawa interaction with $\lambda\ll R$ is preserved
in the presence of these layers. We first note
that the plate thickness ($D_1=3.5\,\mu$m) is much larger
than the typical interaction range of the Yukawa interaction
under consideration ($\lambda=0.1\,\mu$m).
Because of this, the term $\exp(-D_1/\lambda)$ in Eq.~(\ref{eq16})
is negligibly small as compared with unity and can be omitted.
The energy per unit area in a plate-plate configuration and
the Yukawa force in a sphere-plate configuration can then be
simplified to the form
\begin{eqnarray}
&&
E_{\rm Yu}(a)=-2\pi G\rho_1\rho_2\alpha\lambda^3
{\rm e}^{-a/\lambda},
\label{eq17} \\
&&
F_{z,{\rm Yu}}^{sp}(a)=-4\pi^2 G\rho_1\rho_2\alpha\lambda^3
R{\rm e}^{-a/\lambda}.
\nonumber
\end{eqnarray}

Using Eq.~(\ref{eq17}) it is a simple matter to take into
consideration thin layers covering the test bodies in the
experiments of Refs.~\cite{20,21,22,23}. As an example, let us
consider the configuration of two homogeneous plates with
the densities $\rho_1$ (the lower plate) and $\rho_2$
(the upper plate) at a separation $a+\Delta_1^{\!\prime}$,
and assume that the lower plate is covered with an additional
layer of thickness $\Delta_1^{\!\prime}$ and density
$\rho_1^{\prime}$. We then apply the first equality in
Eq.~(\ref{eq17}) to two thick plates with densities $\rho_1^{\prime}$
and $\rho_2$ at a separation $a$ and subtract from the
resulting expression the Yukawa energy per unit area between two
thick plates with densities $\rho_1^{\prime}$
and $\rho_2$ at a separation $a+\Delta_1^{\!\prime}$.
This gives us the Yukawa energy per unit area between
the thin layer of thickness $\Delta_1^{\!\prime}$ and density
$\rho_1^{\prime}$ spaced $a$ apart from the thick upper plate
of density $\rho_2$
\begin{equation}
E_{\rm Yu}^{\Delta_1^{\!\prime}}(a)=-2\pi G\rho_1^{\prime}
\rho_2\alpha\lambda^3{\rm e}^{-a/\lambda}
\bigl(1-{\rm e}^{-\Delta_1^{\!\prime}/\lambda}\bigr).
\label{eq18}
\end{equation}
\noindent
We next combine Eq.~(\ref{eq18}) with the first equality in
Eq.~(\ref{eq17}) written for the Yukawa energy between two
thick plates of densities $\rho_1$ and $\rho_2$
at a separation $a+\Delta_1^{\!\prime}$
and obtain the desired result for the two plates one of which
is covered with a thin layer
\begin{equation}
E_{\rm Yu}(a)=-2\pi G\rho_2\alpha\lambda^3
{\rm e}^{-a/\lambda}\bigr[\rho_1^{\prime}-
(\rho_1^{\prime}-\rho_1){\rm
e}^{-\Delta_1^{\!\prime}/\lambda}\bigr].
\label{eq19}
\end{equation}
\noindent
By repeating the same procedure for each of the two thin layers
covering the lower and upper plate one arrives at
\begin{eqnarray}
&&
E_{\rm Yu}^{\Delta}(a)=-2\pi G\alpha\lambda^3
{\rm e}^{-a/\lambda}
\label{eq20} \\
&&~~
\times
\bigr[\rho_1^{\prime\prime}-
(\rho_1^{\prime\prime}-\rho_1^{\prime}){\rm
e}^{-\Delta_1^{\!\prime\prime}/\lambda}-
(\rho_1^{\prime}-\rho_1){\rm
e}^{-(\Delta_1^{\!\prime\prime}+\Delta_1^{\!\prime})/\lambda}
\bigr]
\nonumber \\
&&~~
\times
\bigr[\rho_2^{\prime\prime}-
(\rho_2^{\prime\prime}-\rho_2^{\prime}){\rm
e}^{-\Delta_2^{\!\prime\prime}/\lambda}-
(\rho_2^{\prime}-\rho_2){\rm
e}^{-(\Delta_2^{\!\prime\prime}+\Delta_2^{\!\prime})/\lambda}
\bigr].
\nonumber
\end{eqnarray}

The vertical component of the Yukawa force given by the
second equality in Eq.~(\ref{eq17}) can also be easily
rewritten for the case when both the sphere and the
plate are covered with two thin layers
\begin{eqnarray}
&&
F_{z,{\rm Yu}}^{sp,\Delta}(a)=-4\pi^2 G\alpha\lambda^3
{\rm e}^{-a/\lambda}
\label{eq21} \\
&&~~
\times
\bigr[\rho_1^{\prime\prime}-
(\rho_1^{\prime\prime}-\rho_1^{\prime}){\rm
e}^{-\Delta_1^{\!\prime\prime}/\lambda}-
(\rho_1^{\prime}-\rho_1){\rm
e}^{-(\Delta_1^{\!\prime\prime}+\Delta_1^{\!\prime})/\lambda}
\bigr]
\nonumber \\
&&~~
\times
\bigr[R\rho_2^{\prime\prime}-(R-\Delta_2^{\!\prime\prime})
(\rho_2^{\prime\prime}-\rho_2^{\prime}){\rm
e}^{-\Delta_2^{\!\prime\prime}/\lambda}-
(R-\Delta_2^{\!\prime\prime}-\Delta_2^{\!\prime})
(\rho_2^{\prime}-\rho_2){\rm
e}^{-(\Delta_2^{\!\prime\prime}+\Delta_2^{\!\prime})/\lambda}
\bigr].
\nonumber
\end{eqnarray}
\noindent
{}From the comparison of Eqs.~(\ref{eq20}) and (\ref{eq21})
it may seem that the equation
\begin{equation}
F_{z,{\rm Yu}}^{sp,\Delta}(a)=2\pi RE_{\rm Yu}^{\Delta}(a)
\label{eq22}
\end{equation}
\noindent
providing a mapping between the configurations of sphere-plate
and plate-plate does not hold for a Yukawa-type interaction.
One should, however, take into account the values of
layer thicknesses
used in the experiments \cite{20,21,22,23}, as compared to sphere
radii. Thus, the thicknesses of the first covering layers vary
from $\Delta_1^{\!\prime}=\Delta_2^{\!\prime}=1\,$nm \cite{20}
to $\Delta_1^{\!\prime}=\Delta_2^{\!\prime}=10\,$nm
\cite{21,22,23}. The thicknesses of the external covering
layers are equal to $\Delta_1^{\!\prime\prime}=
200\,$nm, $\Delta_2^{\!\prime\prime}=203\,$nm \cite{20},
$\Delta_1^{\!\prime\prime}=150\,$nm,
$\Delta_2^{\!\prime\prime}=200\,$nm \cite{21,22}, and
$\Delta_1^{\!\prime\prime}=210\,$nm,
$\Delta_2^{\!\prime\prime}=180\,$nm \cite{23}.
The spheres used in these experiments have the radii
$R=294.3\,\mu$m \cite{20}, $148.7\,\mu$m \cite{21,22},
and $151\,\mu$m \cite{23}.
If we rewrite Eq.~(\ref{eq21}) in the form
\begin{eqnarray}
&&
F_{z,{\rm Yu}}^{sp,\Delta}(a)=-4\pi^2 G\alpha\lambda^3
{\rm e}^{-a/\lambda}R
\label{eq23} \\
&&~~
\times
\bigr[\rho_1^{\prime\prime}-
(\rho_1^{\prime\prime}-\rho_1^{\prime}){\rm
e}^{-\Delta_1^{\!\prime\prime}/\lambda}-
(\rho_1^{\prime}-\rho_1){\rm
e}^{-(\Delta_1^{\!\prime\prime}+\Delta_1^{\!\prime})/\lambda}
\bigr]
\nonumber \\
&&~~
\times
\left[
\vphantom{\frac{\Delta_2^{\!\prime\prime}}{R}}
\rho_2^{\prime\prime}-
(\rho_2^{\prime\prime}-\rho_2^{\prime}){\rm
e}^{-\Delta_2^{\!\prime\prime}/\lambda}-
(\rho_2^{\prime}-\rho_2){\rm
e}^{-(\Delta_2^{\!\prime\prime}+\Delta_2^{\!\prime})/\lambda}
\right.
\nonumber\\
&&~~~~~~
+\left.\frac{\Delta_2^{\!\prime\prime}}{R}
(\rho_2^{\prime\prime}-\rho_2^{\prime}){\rm
e}^{-\Delta_2^{\!\prime\prime}/\lambda}+
\frac{\Delta_2^{\!\prime\prime}+\Delta_2^{\!\prime}}{R}
(\rho_2^{\prime}-\rho_2){\rm
e}^{-(\Delta_2^{\!\prime\prime}+\Delta_2^{\!\prime})/\lambda}
\right],
\nonumber
\end{eqnarray}
\noindent
it is easily seen that in all experiments under consideration
\cite{20,21,22,23} the last two terms on the right-hand side
of Eq.~(\ref{eq23}) are negligibly small, and can be discarded due
to the inequalities
\begin{equation}
\frac{\Delta_2^{\!\prime\prime}}{R}\ll 1,\qquad
\frac{\Delta_2^{\!\prime\prime}+\Delta_2^{\!\prime}}{R}\ll 1.
\label{eq24}
\end{equation}
\noindent
In so doing one neglects quantities
$\approx 0.7\times 10^{-3}$ \cite{20}, $1.4\times 10^{-3}$ \cite{21,22},
and $1.2\times 10^{-3}$ \cite{23} as compared with unity.
This confirms the validity of Eq.~(\ref{eq22}) in the configurations
of experiments \cite{20,21,22,23} with a very high precision
(an error of about 0.1\%). This is the same level of precision
with which the PFA describes the Casimir force in a sphere-plate
configuration (see Sec.~I).

\section{Conclusions and discussion}

In the foregoing we have applied the proximity force approximation
to gravitational and Yukawa forces. This approximate method
is widely used to calculate the Casimir force in configurations
with curved boundaries, but was not considered up to now
in application to gravitation or Yukawa-type interactions.
This allowed some doubts to be raised \cite{32}
 concerning the suitability
of the PFA to slowly decreasing forces of a volumetric
character, and to thus question the validity of constraints obtained
on such forces from dynamic measurements of the Casimir
force \cite{20,21,22,23}.

We have shown that care is required in the application of the
PFA to gravitational forces. Thus the most general
formulation of the PFA, Eq.~(\ref{eq2}), is quite applicable for the
calculation of the gravitational force in  a sphere-plate
configuration while the simplified formulation, Eq.~(\ref{eq3}), is
not. This situation is not unique. For instance, the simplified
formulation is not applicable for the calculation of the
Casimir force in a configuration of a thin transparent
dielectric lens above a plate. At the same time, the most general
formulation of the PFA works well in this case. Another
example is the calculation of the electrostatic force
acting between a plane and a sphere with local deviations
from perfect spherical shape. Here again the most general formulation
of the PFA works well \cite{33}, whereas the simplified formulation
fails to provide a correct theoretical description of the
electric force. By extension the PFA will also apply to inverse
power laws $\sim 1/r^n$ ($n>2$).

For Yukawa-type forces we have demonstrated the applicability
of both formulations of the PFA in a sphere-plate configuration
under the condition that the interaction range
$\lambda$ is much smaller than
the sphere radius. We have also shown that the PFA remains valid
in the experimental configurations of experiments on
the indirect dynamic measurement of the Casimir pressure between
two parallel plates in the configuration of a sphere and
a plate covered with additional thin layers. This allows the
mapping of Yukawa interactions in a sphere-plate and a plate-plate
configurations and confirms the validity of constraints on
non-Newtonian gravity obtained in Refs.~\cite{20,21,22,23}.

\section*{Acknowledgments}

R.S.D.~acknowledges NSF support through Grants No.~CCF-0508239
and PHY-0701636, and from the Nanoscale Imaging Center at IUPUI.
E.F. was supported in part by DOE under Grant No.~DE-76ER071428.
G.L.K. and V.M.M. were partially supported by
Deutsche Forschungsgemeinschaft, Grant No.~436\,RUS\,113/789/0--4
and by Purdue University.


\begin{figure}[b]
\vspace*{-12cm}
\centerline{\hspace*{7cm}
\includegraphics{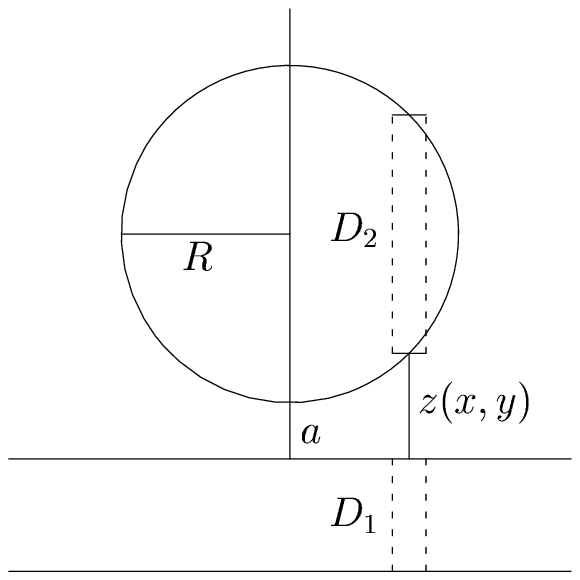}
}
\vspace*{-8cm}
\caption{Configuration of a sphere of radius $R$
spaced at the height $a$ above a plate of thickness $D_1$
(see text for further discussion).
}
\end{figure}

\end{document}